\documentclass[aps,prd,reprint,superscriptaddress,nofootinbib,preprintnumbers]{revtex4-2}

\usepackage{comment}
\usepackage[utf8]{inputenc} 
\usepackage{graphicx,overpic,mathtools}
\usepackage{amsthm,amsmath,amssymb,hyperref,mathrsfs}
\usepackage{braket,bm,bbm,setspace}
\usepackage[normalem]{ulem} 
\usepackage{physics}
\usepackage[makeroom]{cancel}
\hypersetup{
	colorlinks=true,
	linkcolor=blue,
	filecolor=magenta,      
	urlcolor=cyan,
}
\usepackage[usenames,dvipsnames]{xcolor}

\newcommand{\bfk}{\textbf{k}}
\newcommand{\bfx}{\textbf{x}}
\newcommand{\omin}{\omega_{\bfk}^{\text{in}}}
\newcommand{\omout}{\omega_{\bfk}^{\text{out}}}

\begin{document}

\title{No black holes from light}

\author{\'Alvaro \'Alvarez-Dom\'inguez}
\email{alvalv04@ucm.es}
\affiliation{Departamento de F\'{\i}sica Te\'orica and IPARCOS, Universidad Complutense de Madrid, Plaza de las Ciencias 1, 28040 Madrid, Spain}   

\author{Luis J. Garay} 
\email{luisj.garay@ucm.es}
\affiliation{Departamento de F\'{\i}sica Te\'orica and IPARCOS, Universidad Complutense de Madrid, Plaza de las Ciencias 1, 28040 Madrid, Spain}

\author{Eduardo Mart\'in-Mart\'inez}
\email{emartinmartinez@uwaterloo.ca}
\affiliation{Department of Applied Mathematics, University of Waterloo, Waterloo, Ontario, N2L 3G1, Canada}
\affiliation{Institute for Quantum Computing, University of Waterloo, Waterloo, Ontario, N2L 3G1, Canada}
\affiliation{Perimeter Institute for Theoretical Physics, Waterloo, Ontario, N2L 2Y5, Canada}

\author{Jos\'e Polo-G\'omez}
\email{jpologomez@uwaterloo.ca}
\affiliation{Department of Applied Mathematics, University of Waterloo, Waterloo, Ontario, N2L 3G1, Canada}
\affiliation{Institute for Quantum Computing, University of Waterloo, Waterloo, Ontario, N2L 3G1, Canada}
\affiliation{Perimeter Institute for Theoretical Physics, Waterloo, Ontario, N2L 2Y5, Canada}

\begin{abstract}
We show that it is not possible to concentrate enough light to precipitate the formation of an event horizon. We argue that the dissipative quantum effects coming from the self-interaction of light (such as vacuum polarization) are enough to prevent any meaningful buildup of energy that could create a black hole in any realistic scenario.
\end{abstract}
\preprint{IPARCOS-UCM-24-026}

\maketitle

\noindent\textit{\textbf{Introduction.---}} One of the consequences of the fact that energy---and not mass---is the one responsible for the curvature of spacetime is the \textit{a priori} possibility of having massless fields being held together by gravity. These exotic structures (known as \textit{geons}) were first considered by Wheeler~\cite{Wheeler1955,Misner1957,Wheeler1981} for electromagnetic fields. The cases of the (almost massless) neutrinos~\cite{Brill1957} and the gravitational field itself~\cite{Brill1964,Anderson1997} were subsequently studied. These objects are found to be unstable under perturbations~\cite{Perry1999}, leading to either a ``leakage'' of the massless field~\cite{Wheeler1955} or its collapse into a black hole~\cite{Gundlach2003}. In this context, the term \textit{kugelblitz} (German for ``ball lightning'') has become popular as a way to refer to any hypothetical black hole formed by the gravitational collapse of electromagnetic radiation~\cite{YoutubeandWikipedia}.

\textit{Kugelblitze} are allowed by general relativity: there are exact solutions to Einstein-Maxwell equations describing black holes generated by the collapse of electromagnetic energy~\cite{Robinson1962,Senovilla2015}. Kugelblitze have been studied in the context of the cosmic censorship hypothesis~\cite{Lemos1992,Lemos1999,Senovilla2015}, the evaporation of white holes~\cite{Senovilla2015}, dark matter~\cite{Guiot2020}, and have even been proposed as the engine of a really speculative option for interstellar travel~\cite{Crane2009,Crane2011,Lee2015}. However, none of these works take into account quantum effects, which should play an important role in determining whether a kugelblitz can form or not. This is especially so if we are interested in black holes of small sizes such as the artificial ones required in~\cite{Crane2009,Crane2011,Lee2015}.

The hypothetical formation of a kugelblitz---even one with as little energy as to be just a few orders of magnitude above the Planck length---would involve electromagnetic field strengths larger than the threshold above which particle creation stops being exponentially suppressed~\cite{Sauter1931,Heisenberg1936}. This phenomenon hinders the formation of the kugelblitz, since the created particles can scatter out of the region where the radiation is collapsing, carrying their energy with them. The production of electron-positron pairs in an intense electromagnetic background is dominated by the Schwinger effect~\cite{Sauter1931,Heisenberg1936,Schwinger1951}, and the multiphoton Breit-Wheeler process~\cite{Breit1934,Reiss1962}. The former arises non-perturbatively from the low-frequency contributions of the  electromagnetic field, while the latter is perturbative and involves high-frequency modes. Here, we will use the term ``Schwinger effect" to encompass both phenomena. 

In this work, we show that the dissipation of energy via Schwinger effect alone is enough to prevent the formation of kugelblitze with radii ranging from $10^{-29}$ to $10^8$~m. Specifically, we consider the scenario where an external flux of electromagnetic radiation is being focused on a spherical region until there is enough energy to form a Schwarzschild black hole. However, our analysis takes into account that a significant fraction of the energy leaks out of the region of formation due to the Schwinger effect: electron-positron pairs are created inside the region, accelerated by the existing electric field, and subsequently expelled with ultrarelativistic velocities. Our analysis strongly suggests that the formation of black holes solely from electromagnetic radiation is impossible, either by concentrating light in a hypothetical laboratory setting or in naturally occurring astrophysical phenomena.

\noindent\textit{\textbf{Setup.---}} In order to generate a spherically symmetric kugelblitz, it would be necessary to concentrate a critical amount of energy $\epsilon_{\textsc{bh}} = Rc^4/(2G)$ in a sphere of radius~$R$. To achieve this with focused radiation, we assume a constant influx $f$ of electromagnetic energy into the sphere. However, the rate at which the energy focused inside the sphere grows is limited by the dissipation due to the scattering of the radiation (via, e.g., Schwinger effect~\cite{Sauter1931,Heisenberg1936,Schwinger1951,Breit1934,Karplus1951,Reiss1962}, or multiphoton scattering~\cite{Brodsky1970,Jaccarini1970,Budnev1975}). The change of energy~$\epsilon(t)$ inside the sphere is then governed by
\begin{equation}\label{Eq: ODE for the change of energy}
\dot{\epsilon}(t) = 4 \pi R^2 f - D(t),
\end{equation}
where $D(t)$ is the dissipation rate. As a lower bound on the scattered energy, we only consider the Schwinger effect, since, as we will see, to reach energies close to $\epsilon_\textsc{bh}$ solely with electromagnetic energy, we will need electric field strengths above the so-called Schwinger limit \cite{Sauter1931,Heisenberg1936}, \mbox{$1.3 \cdot 10^{18} \text{ V/m}$}. Estimating $D(t)$ is challenging, since the Schwinger effect is non-Markovian~\cite{Kluger1998,Schmidt1999}, while Eq.~\eqref{Eq: ODE for the change of energy} assumes Markovianity. The non-Markovianity of the Schwinger effect is due to the backreaction of the produced pairs on the ability of the background electric field to produce more pairs. 

The production of particles is restricted to the sphere of radius~$R$ where the electromagnetic energy is confined. These particles are scattered in all directions and eventually leave the sphere in some average ``exiting'' time~$\tau_\text{x}$, after which they stop influencing the pair creation. Hence, the fermion production in the sphere after that time can be considered to be ``reset''. Another way of understanding this approximation is to consider that the correlations between pairs of fermions produced in the sphere at different instants $t_1$ and $t_2$ are negligible whenever $|t_1-t_2| \gg \tau_\text{x}$. Thus, as long as $\tau_\text{x}$ is much smaller than the timescale $T$ of the formation of the kugelblitz, it is safe to describe the process in a coarse-grained way that relegates the memory effects to the timescales below $\tau_\text{x}$. The continuous process is hence discretized into a sequence of non-Markovian processes of typical duration~$\tau_\text{x}$, and because $\tau_\text{x} \ll T$, the discrete evolution can be approximated by a continuous one. We will see that this is indeed a good approximation, as we can estimate $\tau_\text{x}$ to be half the light-crossing time of the sphere, $R/c$, and the timescales predicted by Eq.~\eqref{Eq: ODE for the change of energy} for the formation of a kugelblitz are consistent with $R \ll cT$. 

At every instant $t$, we model this process by considering an electric field pulse of maximum strength $E(t)$ that is switched on and off adiabatically, is homogenenous in the sphere, and stays on for a characteristic time~$\tau_{\text{x}}$. Adiabaticity ensures that the tails of the electric field profile contribute negligibly to the particle production~\cite{Adorno2018,Alvarez2023,Ilderton2022,Diez2023}. To compute the energy density carried by the particle-antiparticle pairs, we resort to the commonplace adiabatic regularization and renormalization of the stress-energy tensor, following~\cite{BeltranPalau2020}. To make the calculation concrete, let us consider an adiabatic pulse given by~\cite{Sauter1931,Poschl1933} 
\begin{equation} \label{eq:EPT(t)}
    E_t(\tau)=E(t)/\cosh^2(\tau/\tau_\text{x}).
\end{equation}
$E_{t}(\tau)$ reaches its maximum amplitude~$E(t)$ at~$\tau=0$, vanishes asymptotically for $\tau\rightarrow \pm \infty$, and has a characteristic duration $\tau_\text{x}$. For our purposes we focus on the regime
\begin{equation}\label{Eq: approx}
    eE(t)\hbar \gg m^2 c^3 \quad \text{and} \quad eE(t)\tau_\text{x}^2 c \gg \hbar,
\end{equation}
where $e$ is the electron charge and $m$ is its mass. In this regime, we can show that 
\begin{equation}\label{Eq: approx energy density particle production}
    \lim_{\tau\rightarrow \infty} \langle\hat{T}^{00}(\tau,\bfx)\rangle_{\text{ren}}\approx \frac{2 e^3}{3\pi^3\hbar^2}\tau_\text{x}^2 E(t)^3.
\end{equation}
The details are well known~\cite{Suen1987backreaction,Suen1987reheating,Cooper1989,Kluger1991,Kluger1992,Ferreiro2018a,Ferreiro2018b,Ferreiro2019,BeltranPalau2019} and are reviewed in Appendix~\ref{Appendix: Schwinger}. This gives us the energy dissipated via the Schwinger effect at time $t$. Dividing the result in Eq.~\eqref{Eq: approx energy density particle production} by the duration~$\tau_\text{x}$ and multiplying by the volume of the sphere yields the energy dissipation rate
\begin{equation}\label{Eq: estimated Schwinger dissipation}
D(t) \approx  \frac{8 e^3}{9\pi^2\hbar^2} \tau_\text{x} [R E(t)]^3 . 
\end{equation}
Since the electromagnetic energy in the sphere is \mbox{$\epsilon(t)=(4\pi R^3/3) \varepsilon_0 E(t)^2/2$}, we can rewrite Eq.~\eqref{Eq: ODE for the change of energy} as a first order differential equation for $E(t)$:
\begin{equation}\label{Eq: model}
\varepsilon_0 E(t) \dot{E}(t) = \frac{3}{R}f - \frac{2e^3\tau_\text{x}}{3\pi^3\hbar^2} E(t)^3.
\end{equation}
This equation has a fixed point
\begin{equation}\label{Eq: E infinity}
    E_\infty=\bigg( \frac{9 \pi^3 \hbar^2 f}{2 e^3 R \tau_\text{x}} \bigg)^{1/3},
\end{equation}
and all its solutions are monotonic and convergent to the fixed point as $t\rightarrow \infty$. Because the electric field needs to build up for the kugelblitz to form dynamically, the monotonicity of the solutions implies that, for the kugelblitz to be viable, the fixed point $E_\infty$ must be above the electric field $E_\textsc{bh}$ required to form an electromagnetic black hole, i.e., 
\begin{align}\label{Eq: Electric field BH}
E_\infty > E_\textsc{bh} & = \sqrt{\frac{3}{4\pi R^3} \frac{2\epsilon_\textsc{bh}}{ \varepsilon_0 }} = \sqrt{\frac{3c^4}{4 \pi \varepsilon_0 G}} \frac{1}{ R} = \frac{\phi}{R},
\end{align}
where $\phi = \sqrt{3c^4 / (4 \pi \varepsilon_0 G)} \sim 10^{27} \text{ V}$. Otherwise the electric field in the sphere would stabilize before reaching the critical value $E_\textsc{bh}$, and the black hole would never form. Notice for reference that the strongest electromagnetic fields in nature are found in magnetars~\cite{Mereghetti2015,Turolla2015,Kaspi2017}, a kind of neutron star. Magnetars display magnetic strengths of $10^{11} \text{ T}$, corresponding to electric strengths of $10^{19} \text{ V/m}$. Meanwhile, the strongest electric field achieved so far in the laboratory is of the order of $10^{15} \text{ V/m}$~\cite{Yoon2021}. 

Since $\tau_\text{x}$ is bounded from below by $R/c$, it can be checked that electric field strengths close to $E_\textsc{bh}$ for \mbox{$10^{-29} \text{ m} \lesssim R \lesssim 10^{8} \text{ m}$} fall well within the regime of approximation given by Eq.~\eqref{Eq: approx}. In this regime, the scattered particles are ultrarelativistic  and we can estimate the exiting time by half the light-crossing time of the sphere, \mbox{$\tau_\text{x} \approx R/c$}. Then, $E_\infty > E_\textsc{bh}$ implies that 
\begin{equation}\label{Eq: power needed}
f R > \frac{2 e^3 \phi^3}{9\pi^3 \hbar^2 c} \sim 10^{83} \; \textrm{W/m}.
\end{equation}
The intensity required to form a laboratory-scale kugelblitz ($R \lesssim 1 \text{ m}$) would be of approximately $10^{83} \text{ W/m}^2$, more than 50 orders of magnitude above state-of-the-art laser pulse intensities, which reach \mbox{$10^{27} \text{ W/m}^2$}~\cite{Yoon2021}. For astrophysical sources, the intensity required is still many orders of magnitude above the highest-intensity sources in the universe, including quasars~\cite{Hopkins2007,Sulentic2014,Wolf2024} and supernovae~\cite{Nicholl2020}. Moreover, from Eq.~\eqref{Eq: power needed}, the total power input must satisfy  
\begin{equation}\label{Eq: total power needed}
4 \pi R^2 f \gtrsim R \cdot (10^{84} \; \textrm{W/m}),
\end{equation}
which is far from the bolometric luminosity of the brightest quasars, $10^{41}$ W~\cite{Hopkins2007,Sulentic2014,Wolf2024}, for any kugelblitz radius above the Planck length. This shows that the formation of a kugelblitz requires energy levels that are not achievable either naturally or artificially. 

\noindent\textit{\textbf{Validity of the results.---}} In order to reach our conclusions, we used an admittedly simple description of the formation of a kugelblitz. Here, we justify that we should not expect significant deviations between the extreme orders of magnitude predicted in our simple setup and those of a more sophisticated one. Let us analyze the main approximations made throughout our argument, namely: \mbox{1) estimating} the Schwinger effect dissipation using the specific electric pulse~\eqref{eq:EPT(t)}, \mbox{2) assuming a Minkowski background,} \mbox{3) considering} a uniform electric field, \mbox{4) estimating} the exiting time $\tau_\text{x}$ by half the light-crossing time of the region of formation, and \mbox{5) modelling} dissipation as a Markovian sequence of short non-Markovian processes.

\noindent\textit{1. Electric field pulse profile.---} As we discussed above, the adiabaticity of the process implies that the tails of the pulse have a negligible impact on the particle production. Any smooth peaked pulse would therefore yield the same order of magnitude of particle production. If the adiabatic approximation is not fulfilled then the dissipation effects are even stronger due to extra particle production~\cite{Adorno2018,Alvarez2023,Ilderton2022,Diez2023}, further hindering the formation of a kugelblitz.

\noindent\textit{2. Assumption of  a Minkowski background.---} One could argue that to analyze the formation of kugelblitze one should work with quantum field theory in a dynamically curved background. However, the results of our analysis reveal that the energy densities that one can realistically reach before and after Schwinger dissipation dominates are not only not forming an event horizon, but are also well within the weak (gravitational) field approximation compatible with a flat spacetime (except for the very late stages, which are unreachable anyway), as can be checked by analyzing a characteristic parameter such as the escape velocity (see Appendix~\ref{Appendix: Exiting time}).

\noindent\textit{3. Uniform electric field approximation.---} To evaluate the suitability of this approximation, we need to understand how fermion pair production is affected by the time and space dependence of the electric field, as well as the related presence of a magnetic field. Although the general case is not analytically tractable, many authors have studied how the spatio-temporal dependence of the electromagnetic pulse affects particle production~\cite{Hebenstreit2010,HebenstreitThesis,KohlfurstThesis,Kohlfurst2022,Gies2005,Dunne2005,Dunne2006,Schutzhold2008,HebenstreitThesis,Amat2022,Nikishov1970,Ritus1985,Ruf2009,Bulanov2010multiple,HebenstreitThesis,Jiang2014,Karabali2019}. 

First, there is evidence that the time dependence of the electric field enhances pair production~\cite{Dunne2005,Dunne2006,Schutzhold2008,HebenstreitThesis}, while the space dependence suppresses it~\cite{Gies2005,Dunne2005,Dunne2006,HebenstreitThesis,Amat2022}. The latter, however, is only significant at scales below the ``pair formation length''~\cite{Nikishov1970,Ritus1985,Dunne2005HELagrangians,Dunne2005,Dunne2006,HebenstreitThesis,DiPiazza2012,Gies2016}, \mbox{$\ell = m c^2/(e E)$}. 
In the hypothetical formation of a kugelblitz of radius $R$, the electric field would have to get increasingly closer to $E_\textsc{bh}$, for which the associated pair formation length would be
\begin{equation}
\frac{\ell_\textsc{bh}}{R} = \frac{m c^2}{e E_\textsc{bh} R} = \frac{mc^2}{e \phi} \sim 10^{-22}. 
\end{equation}
Even if we managed to devise a setup for which the Schwinger effect was suppressed for weaker electric fields, suppressing it until the formation of the black hole would require inhomogeneities with typical lengthscales of the order of $\ell_\textsc{bh}$ or below. In laboratory setups, where $R \lesssim 1 \text{ m}$, reaching this regime would require radiation with wavelengths of the order of $10^{-22} \text{ m}$ or below, which is more than ten orders of magnitude below the current shortest laser wavelength~\cite{Yoneda2015}---in fact, $10^{-21} \text{ m}$ is the order of magnitude of the shortest wavelength ever measured for radiation coming from an astrophysical source~\cite{Photons100TeV,PhotonsPeV,Kar2022}. For $\ell_\textsc{bh}$ to approach, for instance, the wavelengths of $\gamma$-ray bursts~\cite{Klebesadel1973,Kouveliotou1993,Bonnell1996,Gendre2013} ($10^{-12}$ m or below~\cite{Ohmori2019}), we would require \mbox{$R \gtrsim 10^9 \,\textrm{m}$}. Outside this regime, the suppression of the Schwinger effect due to spatial inhomogeneities of the electric field is negligible. 

Regarding the magnetic field that would unavoidably exist in a dynamical scenario, its presence increases pair production~\cite{Nikishov1970,Dunne2005HELagrangians,Harko2006,Karabali2019}, and this effect is increased by curvature and strong gravitational fields~\cite{Calucci1999,DiPiazza2002,DiPiazza2006,Karabali2019,Haryanto2023}. While it is true that a single plane electromagnetic wave cannot lead to pair production~\cite{Dunne2005HELagrangians}, the formation of a kugelblitz would require focusing radiation, and possibly multiple sources. This makes things even worse for the formation of a kugelblitz: it has been shown that both focused radiation~\cite{Narozhny2004,Bulanov2005,Narozhny2000} and light from multiple sources~\cite{Brown1964,Brezin1970,Fried2001,Dunne2009,Bulanov2010multiple,Bulanov2010,Mocken2010} are more efficient at pair production than a constant electric field.

Overall, the approximation by a uniform electric field should lead to an underestimation of the dissipation via the Schwinger effect, since the neglected effects either enhance pair production, or are irrelevant in the regimes where a kugelblitz could form~\cite{Ritus1985,Dittrich2000QuantumVacuum,DiPiazza2012}.

\noindent\textit{4. Estimation of the exiting time.---} Notice that to say that $\tau_\text{x} \approx R/c$, we neglected the gravitational attraction that the confined radiation exerts on the scattered fermions. However, in the regime of approximation where the electric field is strong enough to produce particle-antiparticle pairs, the fermions produced by the field quickly become ultrarelativistic, reaching significantly larger velocities than those required to escape the collapsing region (see the straightforward calculation in Appendix~\ref{Appendix: Exiting time}). We could even consider the contrived scenario where one is able to use some external mechanism to confine the fermions. In this situation, most of the energy carried by the particle-antiparticle pairs would still be dissipated away in the form of bremsstrahlung~\cite{Jackson1999classical} produced during their extreme deceleration (see Appendix~\ref{Appendix: Exiting time} for the computational details).

\noindent\textit{5. Markovian approximation.---} We described the dissipation via Schwinger effect as a sequence of independent non-Markovian dissipation processes of negligibly small duration as compared with the time $T$ of formation of the kugelblitz, i.e., $\tau_\text{x} \ll T$. Hence, it is necessary to examine whether this is actually the case. We can bound $T$ from below by the time it would take to form the kugelblitz in the absence of dissipation:
\begin{equation}
T \gtrsim \frac{\epsilon_\textsc{bh}}{4 \pi R^2 f} = \frac{c^4}{8 \pi G R f}.
\end{equation}
On the other hand, we have seen that $\tau_\text{x} \approx R/c$, and therefore
\begin{equation}
\frac{\tau_\text{x}}{T} \lesssim 4 \pi R^2 f \frac{2 G}{c^5}\sim  \frac{4\pi R^2 f}{10^{52}\text{ W} } .
\end{equation}
Hence, for $\tau_\text{x}$ to be non-negligible in comparison with~$T$, the total power input needs to be \mbox{$4\pi R^2 f \sim   10^{51} \text{ W}$}. This lower bound is already ten orders of magnitude above the power output of quasars~\cite{Hopkins2007,Sulentic2014,Wolf2024}. 

Finally, in order to use Eq.~\eqref{Eq: approx energy density particle production} to obtain the rate of dissipation via Schwinger effect, we need to confirm that we fall into the regime of approximation where it is valid, given by Eq.~\eqref{Eq: approx}. The first inequality of Eq.~\eqref{Eq: approx} simply requires that the electric field strength is much larger than the Schwinger limit, \mbox{$1.3 \cdot 10^{18} \text{ V/m}$}, above which pair production takes place. The second inequality can be rewritten as
\begin{equation}
E(t) \tau_\text{x}^2 c^2 \approx E(t) R^2 \gg \frac{\hbar c}{e} \sim 10^{-7} \; \textrm{V m}.
\end{equation}
However, from Eq.~\eqref{Eq: Electric field BH}, \mbox{$E_\textsc{bh} R = \phi \sim 10^{27} \text{ V}$}. Thus, both inequalities are satisfied by $E_\textsc{bh}$ whenever \mbox{$10^{-29} \text{ m} \lesssim R \lesssim 10^{8} \text{ m}$}, and therefore need to be satisfied by $E(t)$ from some instant forward, since it needs to approach $E_\textsc{bh}$ for the kugelblitz to form.

On one hand, $10^{-29} \text{ m}$ is more than 10 orders of magnitude below the smallest focus spot size achieved for a laser~\cite{Self1983,Siegman1986Lasers}, and it is close enough to the Planck length to consider it outside of any naturally occurring phenomena. On the other hand, the amount of energy necessary to form a black hole with $R \gtrsim 10^9 \text{ m}$ is approximately $10^{53} \text{ J}$, which is the energy output of a bright quasar for over $10^4 \text{ years}$~\cite{Hopkins2007,Sulentic2014,Wolf2024}. Nevertheless, black holes of these sizes and larger do exist. These are the so-called supermassive black holes (with masses $M \gtrsim 10^{6} M_{\odot}$)~\cite{Kormendy1995,Ferrarese2005,Volonteri2010,Kormendy2013}. However, the proposed mechanisms for their formation involve the merger of smaller black holes and/or the evolution from an intermediate-mass black hole through the accretion of matter~\cite{Loeb1994,Ebisuzaki2001,Bromm2003,Volonteri2003,Gurkan2004,Volonteri2007,Tanaka2009,Kulier2015,Latif2016,Pacucci2020}, rather than its direct collapse~\cite{Begelman2006}. Thus, the formation of a kugelblitz of these characteristics seems extremely implausible, except for maybe the exceptionally extreme conditions of the very early universe.

\noindent\textit{\textbf{Conclusions.---}} We showed that it is not possible to generate a black hole out of the gravitational collapse of electromagnetic radiation in the range of lengthscales comprised between $10^{-29} \text{ m}$ and $10^8 \text{ m}$. 

To reach this conclusion, we studied the rate at which electromagnetic energy can be focused on a spherical region of a certain radius when a constant inward flux is applied, while part of it is leaked away by the particle-antiparticle pairs created in the process via the Schwinger effect. 

Our analysis indicates that the power needed to form a kugelblitz is tens of orders of magnitude above what can be achieved in any realistic scenario, both in the laboratory and in astrophysical setups. Moreover, we showed that the  approximations incurred in this analysis do not affect the regimes where our conclusions apply. Furthermore, even if one only trusts the estimations of the model to some extent, the predicted orders of magnitude are so vastly unrealistic as to make this study a very compelling argument against the viability of kugelblitze, both artificially or as a naturally occurring phenomenon.

\acknowledgments

The authors would like to thank Mercedes Mart\'in-Benito, Jose A. R. Cembranos, and the rest of the Quantum Fields and Gravity group for their comments and insights. We would also like to thank Elizabeth Winstanley for helpful discussions. AAD and LJG acknowledge  support through the Grant PID2020-118159GB-C44 (funded by MCIN/AEI/10.13039/501100011033). EMM acknowledges support through the Discovery Grant Program of the Natural Sciences and Engineering Research Council of Canada (NSERC). EMM also acknowledges support of his Ontario Early Researcher award. JPG acknowledges the support of a Mike and Ophelia Lazaridis Fellowship, as well as the support of a fellowship from ``La Caixa'' Foundation (ID 100010434, code LCF/BQ/AA20/11820043). Research at Perimeter Institute is supported in part by the Government of Canada through the Department of Innovation, Science and Industry Canada and by the Province of Ontario through the Ministry of Colleges and Universities.

\onecolumngrid
\appendix

\section{Quick review of the Schwinger effect}\label{Appendix: Schwinger}

In this appendix, we calculate the energy density of particles created via Schwinger effect that are scattered out of the sphere where a potential kugelblitz is forming. To achieve this, we compute the energy density of fermions generated by a homogeneous electric field pulse defined in the entire space, which matches the energy density within the region of formation.
As the vacuum expectation value of the energy-momentum tensor involves ultraviolet divergences, we will resort to its adiabatic regularization and renormalization. The adiabatic regularization used in gravitational scenarios in~\cite{Suen1987backreaction,Suen1987reheating} can be readapted in the presence of a homogeneous time-dependent electric field, both in Minkowski~\cite{Cooper1989,Kluger1991,Kluger1992} and Friedmann-Lema\^{i}tre-Robertson-Walker~\cite{Ferreiro2018a,Ferreiro2018b,Ferreiro2019,BeltranPalau2020} spacetimes. We will mostly follow the notation and procedure of Ref.~\cite{BeltranPalau2020}, which deals with Dirac fields in (3+1)-dimensions. We will use natural units \mbox{$\hbar = c = 1$}.

Let us consider a Dirac field~$\psi(\tau,\bfx)$ in Minkowski spacetime coupled to an external electromagnetic potential~$A_{\mu}(\tau,\bfx)$. The dynamics of the fermionic field is governed by the Dirac equation
\begin{equation} \label{eq:Dirac}
    \left[i\gamma^{\mu}\left(\partial_{\mu}+ieA_{\mu}\right)-m\right]\psi=0,
\end{equation}
where~$\gamma^{\mu}$ are the usual Dirac gamma matrices. For a homogeneous time-dependent electric field $\textbf{E}(\tau)$, the temporal gauge $A_{\mu}(\tau,\bfx)=(0,\textbf{A}(\tau))$ is preferred, as it explicitly preserves homogeneity in the equations of motion. We fix this gauge, as well as the direction of the electric field so that \mbox{$\textbf{E}(\tau)=-\dot{A}(\tau)\textbf{e}_3$}. In order to proceed with the adiabatic regularization, it will be useful to work with the unitarily transformed field $\psi^{\prime}(\tau,\bfx)=U\psi(\tau,\bfx)$, with $U=\gamma^0(I-\gamma^3)/\sqrt{2}$, leading to the following reformulation of the Dirac equation~\eqref{eq:Dirac}:
\begin{equation} \label{eq:NewDirac}
    \left[ \gamma^0\partial_0-\gamma^1\partial_1-\gamma^2\partial_2-\partial_3-ieA(\tau)-im\gamma^3 \right]\psi^{\prime}(\tau,\bfx)=0.
\end{equation}

The quantum field operator 
\begin{equation}
\label{eq:QuantumOperator}
    \hat{\psi}^{\prime}(\tau,\bfx)\!=\!\!\!\sum_{\lambda=\pm 1}\!\int\!\! \frac{d\bfk}{(2\pi)^{\frac{3}{2}}}\! \left[ \hat{c}_{\bfk\lambda}u_{\bfk\lambda}(\tau)e^{i\bfk\cdot\bfx}
    + \hat{d}_{\bfk\lambda}^{\dagger}v_{\bfk\lambda}(\tau)e^{-i\bfk\cdot\bfx} \right]
\end{equation}
is defined by expanding the classical field in an orthonormal basis of Fourier spinor solutions of Eq.~\eqref{eq:NewDirac}, $\left\{u_{\bfk\lambda}(\tau)e^{i\bfk\cdot\bfx},v_{\bfk\lambda}(\tau)e^{-i\bfk\cdot\bfx}\right\}_{\bfk,\lambda}$, and  promoting the linear coefficients to annihilation and creation operators~$\hat{c}_{\bfk\lambda}$ and~$\hat{d}_{\bfk\lambda}^{\dagger}$, respectively. These operators satisfy the usual canonical anticommutation relations 
\begin{equation} \label{eq:Anticommutators}
    \{\hat{c}_{\bfk\lambda},\hat{c}_{\bfk'\lambda'}^{\dagger}\}=\{\hat{d}_{\bfk\lambda},\hat{d}_{\bfk'\lambda'}^{\dagger}\}=\delta(\bfk-\bfk')\delta_{\lambda\lambda'},
\end{equation}
and all other anticommutators vanish. The vacuum $\ket{0}$ of the quantized field is then defined as the state annihilated by all the annihilation operators; i.e., $\hat{c}_{\bfk\lambda}\ket{0}=\hat{d}_{\bfk\lambda}\ket{0}=0$, for all values of~$\bfk$ and~$\lambda$. 

We can express a generic basis of orthonormal spinors as
\begin{equation}
    \label{eq:Spinors}
    u_{\bfk\lambda}(\tau)=\mqty(h_{\bfk}^{I}(\tau)\eta_{\lambda}(\bfk) \\h_{\bfk}^{II}(\tau)\lambda\eta_{\lambda}(\bfk)), \qquad v_{\bfk\lambda}(\tau)=\mqty(-h_{-\bfk}^{II*}(\tau)\eta_{-\lambda}(-\bfk) \\-h_{-\bfk}^{I*}(\tau)\lambda\eta_{-\lambda}(-\bfk)),
\end{equation}
where
\begin{equation}
    \label{eq:2DSpinors}
    \eta_{+1}(\bfk)=C \mqty(\kappa+m \\k_1+ik_2), \qquad \eta_{-1}(\bfk)=C \mqty(-k_1+ik_2 \\ \kappa+m)
\end{equation}
form an orthonormal basis of 2-dimensional spinors, and we have defined \mbox{$\bfk=(k_1,k_2,k_3)$}, \mbox{$\bfk_\perp=(k_1,k_2,0)$}, \mbox{$\kappa=\sqrt{k_\perp^2+m^2}$}, and \mbox{$C=1/\sqrt{2\kappa(\kappa+m)}$}. The time-dependent functions $h_{\bfk}^{I}(\tau)$ and $h_{\bfk}^{II}(\tau)$ parametrize the freedom in the choice of the basis and, as dictated by Eq.~\eqref{eq:NewDirac}, they must be solutions to the equations of motion
\begin{align}
    \label{eq:hk}
    \dot{h}_{\bfk}^{I}(\tau)&-i[k_3+eA(\tau)]h_{\bfk}^{I}(\tau)-i\kappa h_{\bfk}^{II}(\tau)=0, \nonumber \\
    \dot{h}_{\bfk}^{II}(\tau)&+i[k_3+eA(\tau)]h_{\bfk}^{II}(\tau)-i\kappa h_{\bfk}^{I}(\tau)=0.
\end{align}
From the normalization of the spinors \eqref{eq:Spinors}, $|h_{\bfk}^{I}(\tau)|^2+|h_{\bfk}^{II}(\tau)|^2=1$.

The energy-momentum tensor of the Dirac field $\psi(\tau,\bfx)$ is given by
\begin{equation}
    T^{\mu\nu}=\frac{i}{4}\left[ \bar{\psi}\gamma^{\mu}(\partial^{\nu}\psi) + \bar{\psi}\gamma^{\nu}(\partial^{\mu}\psi) - (\partial^{\mu}\bar{\psi})\gamma^{\nu}\psi -(\partial^{\nu}\bar{\psi})\gamma^{\mu}\psi  +2ie\left( A^{\mu}\bar{\psi}\gamma^{\nu}\psi + A^{\nu}\bar{\psi}\gamma^{\mu}\psi \right) \right],
\end{equation}
where $\bar{\psi}(\tau,\bfx)=\psi^{\dagger}(\tau,\bfx)\gamma^0$ denotes the adjoint field. The vacuum expectation value of its time-time component in terms of the  transformed quantum field $\hat{\psi'}(\tau,\bfx)$ can be written (with the use of \eqref{eq:QuantumOperator}--\eqref{eq:2DSpinors}) as
\begin{equation} \label{eq:ExpT}
    \langle\hat{T}^{00}\rangle=\bigg\langle\frac{i}{2}\left[ \hat{\psi}^{\prime\dagger}\left(\partial_\tau \hat{\psi}^{\prime}\right) - \left(\partial_\tau \hat{\psi}^{\prime\dagger}\right) \hat{\psi}^{\prime} \right]\bigg\rangle =-2 \int \frac{d\bfk}{(2\pi)^3} \  \rho_{\bfk}(\tau), \qquad \rho_{\bfk}(\tau)=\Im{h_{\bfk}^{I}(\tau)\dot{h}_{\bfk}^{I*}(\tau)+h_{\bfk}^{II}(\tau)\dot{h}_{\bfk}^{II*}(\tau)},
\end{equation}
which still requires renormalization.

From now on, we consider the analytically solvable Sauter electric potential~\cite{Sauter1931}
\begin{equation}
    A_t(\tau)=-E(t)\tau_\text{x}\tanh{(\tau/\tau_\text{x})},
\end{equation}
which corresponds to the  electric pulse  in Eq.~\eqref{eq:EPT(t)}. In the asymptotic past there is still no electric field on, thus the in-solutions $h_{\bfk}^{I}(\tau)$, $h_{\bfk}^{II}(\tau)$ to Eq.~\eqref{eq:hk} are the particular solutions that asymptotically behave as  Minkowski positive-frequency plane waves, i.e., 
\begin{equation}
    h_{\bfk}^{I/II}(\tau) \sim \pm\sqrt{\frac{\omin \mp [k_3+eE(t)\tau_\text{x}]}{2\omin}}e^{-i\omin \tau} 
    \qquad \text{when}\quad \tau\rightarrow -\infty,
\end{equation}
where $\omin=\sqrt{[k_3+eE(t)\tau_\text{x}]^2+\kappa^2}$. The evolution of these in-solutions can be written in terms of hypergeometric functions \cite{BeltranPalau2019}. In the out-region, when the electric field is asymptotically switched-off, the in-solutions will be a linear combination of positive and negative-frequency out-plane waves:
\begin{align}
\label{eq:hkSauter}
 h_{\bfk}^{I/II}(\tau) \sim  \pm \alpha_{\bfk}\sqrt{\frac{\omout \mp [k_3-eE(t)\tau_\text{x}]}{2\omout}}e^{-i\omout \tau} %\nonumber \\ & 
 +\beta_{\bfk}\sqrt{\frac{\omout \pm [k_3-eE(t)\tau_\text{x}]}{2\omout}}e^{i\omout \tau}
 \qquad \text{when}\quad \tau\rightarrow -\infty,
\end{align}
where $\omout=\sqrt{[k_3-eE(t)\tau_\text{x}]^2+\kappa^2}$. This is a usual Bogoliubov transformation, where $\alpha_{\bfk}$ and $\beta_{\bfk}$ are the Bogoliubov coefficients satisfying $|\alpha_{\bfk}|^2+|\beta_{\bfk}|^2=1$. Note that $\beta_{\bfk}$ quantifies the part of the positive-frequency in-solution that transforms into a negative-frequency out-solution. Thus, $n_{\bfk}=|\beta_{\bfk}|^2$ provides the density number of excitations out of the vacuum due to the application of the external electric field. For the electric pulse profile considered here, it is given by~\cite{BeltranPalau2019}
\begin{equation} \label{eq:nkSauter}
    n_{\bfk}=\frac{\cosh{\left[2\pi eE(t)\tau_\text{x}^2\right]}-\cosh{\left[\pi\left(\omout-\omin\right)\tau_\text{x}\right]}}{2\sinh{\left[\pi\omin\tau_\text{x}\right]}\sinh{\left[\pi\omout\tau_\text{x}\right]}}.
\end{equation}
Substituting Eq.~\eqref{eq:hkSauter} in the energy density per mode (Eq.~\eqref{eq:ExpT}) we obtain in the asymptotic future
\begin{equation} \label{eq:rhofuture}
    \lim_{\tau\rightarrow +\infty} \rho_{\bfk}(\tau) = \left(1-2n_{\bfk}\right)\omout,
\end{equation}
where $1-2n_{\bfk}$ is the usual Pauli-blocking factor. 

To proceed with the adiabatic regularization \cite{Ferreiro2019,BeltranPalau2020} of the expectation value of the time-time component of the quantum energy-momentum tensor \eqref{eq:ExpT}, we compute the energy density $\rho_{\bfk}^{\text{ad}}(\tau)$ for the zeroth-order adiabatic modes\footnote{Here, we assign a zero adiabatic order to the electric potential~$A_t(\tau)$. It is noteworthy that when the background includes both gravitational and electromagnetic contributions, Refs.~\cite{Ferreiro2018a,Ferreiro2018b} state that the adiabatic regularization scheme must be performed treating $A_t(\tau)$ as a variable of first adiabatic order. However, in a pure electromagnetic background in Minkowski spacetime, the adiabatic order of the electromagnetic potential is irrelevant for the computation of physical observables.}
\begin{equation}
    h_{\bfk}^{I/II\text{ad}}(\tau)=\pm\sqrt{\frac{\omega_{\bfk}(\tau)\mp [k_3+eA_t(\tau)]}{2\omega_{\bfk}(\tau)}}e^{-i\int^\tau d\tau' \omega_{\bfk}(\tau') },
\end{equation}
where the time-dependent frequency is given by~$\omega_{\bfk}(\tau)=\sqrt{[k_3+eA_t(\tau)]^2+\kappa^2}$. Up to zeroth-order, this results in~$\rho_{\bfk}^{\textrm{ad}}(\tau)=\omega_{\bfk}(\tau)$. Then, subtracting this contribution, we obtain
\begin{equation}
\label{eq:T00ren}
    \lim_{\tau\rightarrow +\infty} \langle\hat{T}^{00}(\tau,\bfx)\rangle_{\text{ren}}=\lim_{\tau\rightarrow +\infty}\left[\langle\hat{T}^{00}(\tau,\bfx)\rangle - \langle\hat{T}^{00}(\tau,\bfx)\rangle_{\text{ad}} \right]
    =\int \frac{d\bfk}{(2\pi)^3} \ 4 n_{\bfk}\omout.
\end{equation}
The factor of $4$ comes from the contribution of each function $h_{\bfk}^{I}(\tau)$ and $h_{\bfk}^{II}(\tau)$, being each one the same for the two different spins.

In the regime where~$eE(t)\tau_\text{x}^2\gg 1$, as assumed in Eq.~\eqref{Eq: approx}, the particle density~\eqref{eq:nkSauter} for the Sauter-type electric potential yields
\begin{equation}
    n_{\bfk} \approx \exp\left( -\frac{\pi\kappa^2}{eE(t)} \frac{1}{1-k_3^2/[eE(t)\tau_\text{x}]^2} \right)\Theta (eE(t)\tau_\text{x}-|k_3|).
\end{equation}
The negative exponent suppresses the contribution of modes with large~$\kappa$ (with respect to $\sqrt{eE(t)}$) in the integral~\eqref{eq:T00ren}. Thus, when~$eE(t)\tau_\text{x}^2\gg 1$, we can approximate $\omout=\sqrt{[k_3-eE(t)\tau_\text{x}]^2+\kappa^2}\approx eE(t)\tau_\text{x}-k_3$ in the domain of the integral. As a result, we obtain
\begin{equation}
    \lim_{\tau\rightarrow +\infty} \langle\hat{T}^{00}(\tau,\bfx) \rangle_{\textrm{ren}} \approx \frac{1}{\pi^2}\int_{0}^{+\infty} dk_3  \int_{0}^{+\infty}dk_{\perp} k_{\perp} \ [eE(t)\tau_\text{x}-k_3] \exp\left( -\pi\frac{k_{\perp}^2+m^2}{eE(t)} \frac{1}{1-k_3^2/[eE(t)\tau_\text{x}]^2} \right).
\end{equation}
The integration in the tranverse momentum $k_{\perp}=\sqrt{k_1^2+k_2^2}$ is straightforward. After simple parity arguments and the change of variables $u=k_3/[eE(t)\tau_\text{x}]$, it leads to
\begin{equation}\label{Eq: final integral}
    \lim_{\tau\rightarrow +\infty} \langle\hat{T}^{00}(\tau,\bfx)\rangle_{\textrm{ren}} \approx \frac{[eE(t)]^3\tau_\text{x}^2}{\pi^3} \int^1_0 du \ (1-u^2)\exp\left( -\pi\frac{m^2}{eE(t)} \frac{1}{1-u^2} \right).
\end{equation}
The values of~$u$ around~$1$ give negligible contributions to the integral. For other values of~$u$, assuming the strong field regime~$eE(t)\gg m^2$ considered in Eq.~\eqref{Eq: approx}, we can approximate the exponential term as one, resulting in Eq.~\eqref{Eq: approx energy density particle production}. 

Finally, notice that in our setup the particle production does not happen in free space, but inside a bounded region of diameter $2R$. However, since the calculation was performed for a homogeneous scenario, the density of produced particles is also homogeneous. Therefore, the only effect that the bounded region has is the removal of the contribution from the modes with $k_3 \lesssim \pi/R$. After the change of variables preceding Eq.~\eqref{Eq: final integral}, this corresponds to $u \lesssim \pi/[e E(t) \tau_\text{x} R] \approx \pi/[e E(t) \tau_\text{x}^2] \ll 1$, and hence the contribution from this range of frequencies is negligible in comparison with the result of the integral in $u$ of Eq.~\eqref{Eq: final integral}, which is $\mathcal{O}(1)$.

\section{Estimation of the exiting time and impossibility of confinement}\label{Appendix: Exiting time}

In this appendix, we will provide the details behind the estimation of the exiting time as $\tau_\text{x} \approx R/c$, as well as the explicit straightforward calculations that justify why confining the exiting charged particles to reduce the dissipation is not enough to enable the formation of a kugelblitz.

As discussed in the main text, the setup under analysis falls within the regime of approximation given by Eq.~\eqref{Eq: approx}. In this regime, the energy of fermions produced via Schwinger effect (in the centre-of-mass reference frame of the collapsing light) can be approximated by
\begin{equation}\label{Eq: energy of a fermion}
\gamma m c^2 \approx e E \tau_\text{x} c,
\end{equation}
where $\gamma=1/\sqrt{1-\beta^2}$ is the Lorentz factor (with $\bm\beta=\bm v/c$, and $\beta\equiv |\bm\beta|$), and we have omitted the dependence of $E$ on $t$ for the sake of a lighter notation. From this, the velocity of the fermions can be approximated by
\begin{equation}\label{Eq: estimated velocity}
v \approx c \,\sqrt{1 - \bigg( \frac{m c}{e E \tau_\text{x}} \bigg)^{\!\!2}}\approx c,
\end{equation}
since in this regime of approximation $e E \gg m^2 c^3/\hbar$ and $eE \tau_\text{x}^2 \gg \hbar/c$, and multiplying both inequalities yields $(e E \tau_\text{x})^2 \gg m^2 c^2$. This means that fermions produced via Schwinger effect are ultrarelativistic. Hence, for the ensemble of particles produced at random positions and with random directions of movement within the sphere of radius $R$, spherical symmetry implies that the average exiting time is
\begin{equation} \label{Eq: estimation taue}
\tau_\text{x} \approx \frac{R}{v} \approx \frac{R}{c}.
\end{equation}

There are two issues that one could raise to challenge this estimation of the exiting time: 1) the gravitational attraction exerted by the confined radiation on the scattered fermions was neglected, and 2) the possibility of using some additional technique to confine the produced fermion pairs was not considered.

Regarding the first issue it should suffice to check that the estimated fermion velocity $v$ from Eq.~\eqref{Eq: estimated velocity} is much larger than  the necessary velocity $v_\text{esc}$ to escape the region of formation of the kugelblitz. This escape velocity can be easily estimated as
\begin{equation}
v_\text{esc} \approx \sqrt{\frac{2G\epsilon}{R c^2}} = c \,\sqrt{\frac{\epsilon}{\epsilon_\textsc{bh}}} = \frac{c E}{E_\textsc{bh}},  
\end{equation}
where recall that $\epsilon$ is the electromagnetic energy in the region of formation, $\epsilon_\textsc{bh}$ is the threshold electromagnetic energy needed to form the black hole, and $E_\textsc{bh}$ is its associated electric field strength. Then, from Eq.~\eqref{Eq: estimated velocity}, 
\begin{equation}
\frac{v_\text{esc}}{v} \approx \frac{E}{E_\textsc{bh}},
\end{equation}
which means that the escape velocity is only comparable to the velocity of the scattered ultrarelativistic particles when the electric field strength $E$ is comparable with $E_\textsc{bh}$. This implies that the gravitational influence of the radiation on the exiting fermions is negligible except in the very final stages where the formation would be imminently taking place. However, the calculations made in the main text show that it is not realistically possible to reach neither $E_\textsc{bh}$ nor any significant fraction of it. Thus, to arrive at our conclusions it is never necessary to work outside of the regime where we can safely assume that the velocity estimated in Eq.~\eqref{Eq: estimated velocity} is always significantly bigger than the escape velocity of the region where the kugelblitz is forming. 

Regarding the second issue, we can show that confining the scattered fermions does not eliminate the dissipation, it just changes the form in which it happens. This is so because the deceleration required to stop the particles leads to  bremsstrahlung that quickly scatters off the region where the kugelblitz is forming. This radiation, as we will see, carries a significant enough fraction of the energy that the fermions initially had, hence yielding a ``corrected'' estimated dissipation that is still large enough for the conclusions of the main text to hold.

To see that this is the case, first recall that the power radiated by an accelerating charge is given by~\cite{Jackson1999classical}
\begin{equation}\label{Eq: radiated power}
\frac{\text{d} \epsilon_\text{rad}}{\text{d} \tau} = \frac{2}{3} r_e m c \gamma_{\tau}^6 \big[\dot{\bm\beta}_{\tau}^2 - (\bm\beta_{\tau}\times\dot{\bm\beta}_{\tau})^2\big],
\end{equation}
where $\bm\beta_\tau = \bm v(\tau)/c$ is the velocity of the particle at time $\tau$ (normalized by $c$), $\gamma_\tau = 1 / \sqrt{1-\beta_\tau^2}$ is its Lorentz factor, and $r_e = e^2 / (4 \pi \varepsilon_0 m c^2)$ is the classical electron radius. We can write $\bm{\dot{\beta}}_t = \bm{\dot{\beta}}_\tau^{\parallel} + \bm{\dot{\beta}}_\tau^{\perp}$, where $\bm{\dot{\beta}}_\tau^{\parallel}$ and $\bm{\dot{\beta}}_\tau^{\perp}$ are, respectively, the tangent and normal components of the acceleration. First, let us consider the scenario where the charge is \textit{completely} stopped before reaching some radius $R' \geq R$ (we will later analyze the case in which the confinement is achieved by forcing the charge to orbit inside the region of radius~$R'$). Then, the power radiated can be bounded from below neglecting the contribution that comes from the normal component:
\begin{equation}\label{Eq: first bound}
\frac{\text{d} \epsilon_\text{rad}}{\text{d} \tau} = \frac{2}{3} r_e m c \gamma_{\tau}^6 \big[(\dot{\bm\beta}_{\tau}^{\parallel})^2 + (\dot{\bm\beta}_{\tau}^{\perp})^2(1-\bm\beta_\tau^2) \big] \geq \frac{2}{3} r_e m c \gamma_{\tau}^6 (\dot{\bm\beta}_{\tau}^{\parallel})^2=\frac{2}{3} r_e m c \frac{\gamma_\tau^2 \dot\gamma_\tau^2}{\gamma_\tau^2 - 1}.
\end{equation}
We can estimate the energy radiated by the charge after it has been slowed down to a certain $\beta_\text{b} < \beta$, where $\beta$ denotes the velocity that the charge initially had when it left the region of formation of the kugelblitz. Taking that initial time to be $\tau=0$, and denoting with $\tau_\text{b}$ the time that it takes to decelerate the charge, we can use the Cauchy-Schwarz inequality to obtain a lower bound for the total radiated energy: 
\begin{equation}
    \epsilon_\text{rad} \geq \frac{2r_emc}{3\tau_\text{b}} \bigg( \int_0^{\tau_\text{b}}\text{d}\tau\, \frac{\gamma_\tau\dot\gamma_\tau}{\sqrt{\gamma_\tau^2-1}} \bigg)^{\!\!2} = \frac{2r_e mc}{3\tau_\text{b}} \bigg(\sqrt{\gamma_\text{b}^2-1}-\sqrt{\gamma^2-1}\bigg)^{\!\!2},
\end{equation}
where $\gamma_\text{b} = 1 / \sqrt{1-\beta_\text{b}^2}$ is the Lorentz factor at $\tau = \tau_\text{b}$. Since we are assuming that the charge is confined in some region of radius $R'\geq R$, $\tau_\text{b}$ must be bounded from above by the time it would take for the particle to leave the region of radius $R'$ with a velocity greater or equal than $\beta_\text{b}$, which at the same time can be upper-bounded by $R'/(c\beta_\text{b})$, yielding
\begin{equation}
    \epsilon_\text{rad} \geq \frac{2r_e mc^2}{3R'} \beta_\text{b} \bigg(\sqrt{\gamma_\text{b}^2-1}-\sqrt{\gamma^2-1}\bigg)^{\!\!2} \approx \frac{2r_e mc^2}{3R'} \beta_\text{b} \bigg(\frac{\beta_\text{b}}{\sqrt{1-\beta_\text{b}^2}}-\gamma\bigg)^{\!\!2} ,
\end{equation}
where we used that the charge is initially ultrarelativistic,~$\gamma \gg 1$. Now, to estimate the fraction~$\chi$ of the initial energy of the charge that leaves the region of formation of the kugelblitz, we can bound it from below by only taking into account the energy radiated until~$\tau=\tau_\text{b}$, and assuming that the remaining energy stays with the charge, which itself remains inside the region of radius $R'$. Notice that the charge can stay \textit{anywhere} inside the region of radius $R'$, of which the region of formation of the kugelblitz is only a subregion of radius $R$. This means that the energy of the fermions that have been stopped is ``diluted'' by a factor that accounts for their different volumes, namely $(R/R')^3$. Therefore, the energy dissipated from the region of formation of the kugelblitz is the initial energy carried by the charge, $\gamma m c^2$, minus the energy that remains, $(\gamma m c^2 - \epsilon_\text{rad})$, multiplied by the ``dilution factor'' $(R/R')^3$. From this, the fraction of dissipated energy is
\begin{equation}\label{Eq: fraction that stays}
    \chi = 1 - \frac{R^3}{R'^3}\bigg( 1 - \frac{\epsilon_\text{rad}}{\gamma m c^2} \bigg) \geq 1 - \frac{R^3}{R'^3} \bigg[ 1 - \frac{2r_e}{3\gamma R'}\beta_\text{b}\bigg( \frac{\beta_\text{b}}{\sqrt{1-\beta_\text{b}^2}} - \gamma \bigg)^{\!\!2}  \bigg].
\end{equation}
Notice that to derive this inequality we only required $\beta_\text{b} < \beta$, and therefore Eq.~\eqref{Eq: fraction that stays} actually represents an infinite set of inequalities. Although we could optimize over $\beta_\text{b}$ to obtain the stricter lower bound of $\chi$, for the sake of simplicity we will just use $\beta_\text{b}=1/2$. In this case,
\begin{equation}\label{Eq: fraction particular case}
    \chi \geq 1 - \frac{R^3}{R'^3}\bigg[ 1 - \frac{r_e}{3\gamma R'} \bigg(\frac{1}{\sqrt{3}}-\gamma\bigg)^{\!\!2} \bigg] \approx 1 - \bigg( \frac{R}{R'} \bigg)^{\!\!3} \bigg[  1 - \frac{r_e \gamma}{3 R} \bigg(\frac{R}{R'}\bigg) \bigg].
\end{equation}
For $R/R'>3R/(r_e \gamma)$, Eq.~\eqref{Eq: fraction particular case} implies that $\chi > 1$, which is impossible. What actually happens in this case is that $R'$ is ``too small'', and the exiting charges cannot be stopped before they leave the region of radius $R'$ (since to do that they would need to dissipate more energy than the charged particle has). Therefore, the confinement of the electron-positron pairs can only be attempted for $R'$ such that $R/R'\leq 3R/(r_e \gamma)$, and only then the bound given by Eq.~\eqref{Eq: fraction particular case} applies. Moreover, since $\gamma \approx e E \tau_\text{x}/ (m c) \geq e E R / (m c^2)$,
\begin{equation}
    \frac{r_e \gamma}{3 R} \gtrsim \frac{r_e e}{3 m c^2} E \sim \frac{E}{10^{21} \text{ V/m}}.
\end{equation}
We can therefore distinguish two regimes: 1) $E \gg 10^{21} \text{ V/m}$, in which case $r_e \gamma / (3 R) \gg 1$, and 2) $E \lesssim 10^{21} \text{ V/m}$, when $r_e \gamma / (3 R) \lesssim 1$. In the first case, $r_e \gamma / (3 R) \gg 1$ implies that either $R/R'>3R/(r_e\gamma)$, and the charges cannot be stopped before leaving $R'$, or $R/R' \leq 3R/(r_e\gamma) \ll 1$, and $\chi \geq 1 - (R/R')^3 \sim 1$, i.e., most of the energy leaves the region of formation of the kugelblitz. In the second regime, we can still use that, in order to form a kugelblitz in the range of radii under analysis, the electric field has to be bigger than the Schwinger limit, $E \gtrsim 10^{18} \text{ V/m}$, implying that $r_e\gamma / (3R) \gtrsim 10^{-3}$. In this case,
\begin{equation}
    \chi \geq  1 - \bigg( \frac{R}{R'} \bigg)^{\!\!3} \bigg[  1 - \frac{r_e \gamma}{3 R} \bigg(\frac{R}{R'}\bigg) \bigg] \gtrsim \frac{r_e \gamma}{3 R} \gtrsim 10^{-3}, 
\end{equation}
where in the second inequality we have used that, as a function of $R/R'$, the right hand side of Eq.~\eqref{Eq: fraction particular case} attains its minimum at $(R/R')^* = \min\{1,9R/(4r_e \gamma)\} \approx 1$. This estimation would mean that with this setup the dissipation is, in the best case, three orders of magnitude below the one used originally, given by Eq.~\eqref{Eq: estimated Schwinger dissipation}. However, from Eq.~\eqref{Eq: E infinity} we see that a correction factor of $10^{-3}$ for the dissipation term $D$ would only increase an order of magnitude the value of $E_\infty$. Even in this extremely optimistic scenario, the resulting reduction is far from being enough to close the huge gap between the estimations of the required conditions to form a kugelblitz and what seems realistically achievable. 

Finally, let us analyze the case where the charges are confined in the region of radius $R' \geq R$ not by slowing them down before they leave, but by making them orbit inside the region instead. In that scenario, the main contribution to the radiated energy comes from the normal component of the acceleration. To give a lower bound on the energy radiated by the confined charges, we look at the contribution of the normal component in Eq.~\eqref{Eq: radiated power}, which yields
\begin{equation}\label{Eq: power radiated synchrotron}
    \frac{\text{d}\epsilon_\text{rad}}{\text{d}\tau} \geq \frac{2}{3} r_e m c \gamma_\tau^6 (\dot{\bm\beta}_\tau^\perp)^2 (1 - \bm\beta_\tau^2) \geq \frac{2 r_e m c^3}{3 R'^2} \gamma_\tau^6 \beta_\tau^4 (1-\beta_\tau^2) = \frac{2 r_e m c^3}{3 R'^2} (\gamma_\tau^2-1)^2, 
\end{equation}
where in the last inequality we have used that the radius of the orbit must be less or equal than $R'$, and therefore we have that $|\dot{\bm\beta_\tau^\perp}| \geq \bm\beta_\tau^2 c / R'$. Now, because the energy radiated is energy lost by the particle, we have that $\dot{\epsilon}_\text{rad} = - \dot\gamma_\tau m c^2$, and thus
\begin{equation}\label{Eq: dot gamma}
\frac{\text{d}\gamma_\tau}{\text{d}\tau} \leq - \frac{2 r_e c}{3 R'^2} (\gamma_\tau^2-1)^2.
\end{equation}
In order to compare the power dissipated in this setup with the estimated dissipation $D$ of the main text (cf. Eq.~\eqref{Eq: estimated Schwinger dissipation}), we can compute the fraction~$\chi$ of the initial energy of the charge that still dissipates in this setup over a period of time equivalent to the timescale $\tau_\text{x}$ of the dissipation in the original setup, 
\begin{equation}\label{Eq: fraction synchrotron}
    \chi = 1 - \bigg( \frac{R}{R'} \bigg)^{\!\!3} \frac{\gamma_\text{x} m c^2}{\gamma m c^2},
\end{equation}
where $\gamma_\text{x}$ is the final Lorentz factor. To bound $\gamma_\text{x}$ from above, we can integrate Eq.~\eqref{Eq: dot gamma} from the time $\tau=0$ when the charge left the region of formation of the kugelblitz to $\tau=\tau_\text{x}$, obtaining
\begin{equation}
   \int_{0}^{\tau_\text{x}} \text{d}\tau \frac{\dot\gamma_\tau}{(\gamma_\tau^2-1)^2} = \left[ -\frac{\gamma_\tau}{2(\gamma_\tau^2-1)} + \frac{1}{4}\log\frac{\gamma_\tau+1}{\gamma_\tau-1} \right]_{\gamma_\tau=\gamma}^{\gamma_\tau=\gamma_\text{x}} \leq  -\frac{2 r_e c}{3 R'^2}\tau_\text{x},
\end{equation}
Since $\tau_\text{x} \geq R/c$, and $\gamma_\text{x} < \gamma$, the previous inequality can be rewritten as
\begin{equation}
    \frac{2 r_e R}{3 R'^2} \leq \left[ -\frac{\gamma_\tau}{2(\gamma_\tau^2-1)} + \frac{1}{4}\log\frac{\gamma_\tau+1}{\gamma_\tau-1} \right]_{\gamma_\tau=\gamma_\text{x}}^{\gamma_\tau=\gamma} \leq \frac{1}{2}\bigg( \frac{\gamma_\text{x}}{\gamma_\text{x}^2-1} - \frac{\gamma}{\gamma^2-1}  \bigg), 
\end{equation}
where in the last step we used that $\log[(\gamma_\tau+1)/(\gamma_\tau-1)]$ is a decreasing function of $\gamma_\tau$. From here, and since, as discussed, $\gamma\gg 1$, we get
\begin{equation}
   \frac{\gamma_\text{x}}{\gamma_\text{x}^2-1} - \frac{1}{\gamma} \geq \frac{4 r_e R}{3 R'^2}.
\end{equation}
Solving the corresponding quadratic inequality, we arrive at
\begin{equation}\label{Eq: inequality for gamma x}
    \gamma_\text{x} \lesssim \frac{1}{2}\bigg( \frac{1}{\gamma} + \frac{4 r_e R}{3 R'^2}   \bigg)^{\!\!-1} +\frac{1}{2}\sqrt{1 + \bigg( \frac{1}{\gamma} + \frac{4 r_e R}{3 R'^2}   \bigg)^{\!\!-2}}. 
\end{equation}
If $4r_e R / (3 R'^2) \gtrsim 1$, then Eq.~\eqref{Eq: inequality for gamma x} implies
\begin{equation}
    \gamma_\text{x} \lesssim \frac{1}{2}\bigg(\frac{3 R'^2}{4 r_e R}   \bigg) +\frac{1}{2}\sqrt{1 + \bigg(\frac{3 R'^2}{4 r_e R}   \bigg)^{\!\!2}} \lesssim \frac{1+\sqrt{2}}{2},
\end{equation}
where we used again that $\gamma \gg 1$. In this case, $\gamma_\text{x}/\gamma \ll 1$, and from Eq.~\eqref{Eq: fraction synchrotron} we conclude that $\chi\sim 1$, i.e., most of the energy leaves the region of formation of the kugelblitz. If, on the contrary, $4r_e R / (3 R'^2) \ll 1$, then Eq.~\eqref{Eq: inequality for gamma x} reduces to
\begin{equation}
    \gamma_\text{x} \lesssim \bigg( \frac{1}{\gamma} + \frac{4 r_e R}{3 R'^2}   \bigg)^{\!\!-1},
\end{equation}
yielding
\begin{equation}\label{Eq: fraction synchrotron case 2}
    \chi = 1 - \bigg( \frac{R}{R'} \bigg)^{\!\!3} \frac{\gamma_\text{x} m c^2}{\gamma m c^2} \gtrsim 1 - \bigg(\frac{R}{R'}\bigg)^{\!\!3} \bigg[1 + \frac{4 r_e e}{3 m c^2} E \bigg( \frac{R}{R'} \bigg)^{\!\!2}\bigg]^{\!-1},
\end{equation}
where as before we used that $\gamma \approx e E \tau_\text{x} / (m c) \geq e E R / (m c^2)$. Since, as a function of $R/R'$, the right hand side of Eq.~\eqref{Eq: fraction synchrotron case 2} is minimized at $(R/R')^* = 1$, we can bound $\chi$ from below as
\begin{equation}\label{Eq: fraction bound}
    \chi \geq 1 - \bigg[ 1 + \frac{4 r_e e}{3 m c^2} E \bigg]^{\!-1} \sim 1 - \bigg( 1 + \frac{E}{10^{20} \text{ V/m}} \bigg)^{\!\!-1}.
\end{equation}
Then, as in our first analysis, we can distinguish two regimes: 1) $E \gg 10^{20} \text{ V/m}$, and 2) $E \lesssim 10^{20} \text{ V/m}$. In the first case, the second term of Eq.~\eqref{Eq: fraction bound} becomes negligible, and, again, $\chi \gtrsim 1$, i.e., the dissipation is essentially the same as in the original setup. In the second regime, we can use once more that $E \gtrsim 10^{18} \text{ V/m}$, in which case $\chi \gtrsim 10^{-2}$. However, as we argued before, even this best-case-scenario correction factor of $10^{-2}$ for the dissipated power $D$ does not modify the conclusions reached in the main text, and would not allow any realistic scenario to satisfy the necessary conditions for the formation of a kugelblitz.

\twocolumngrid

\bibliography{BibKugelblitz}

\end{document}